\documentclass[aps,prb,showpacs,twocolumn,amsmath]{revtex4}
\usepackage{graphicx}

\newcommand{\bq}{\begin{equation}}
\newcommand{\eq}{\end{equation}}
\newcommand{\bn}{\begin{eqnarray}}
\newcommand{\en}{\end{eqnarray}}
\newcommand{\bsub}{\begin{subequations}}
\newcommand{\esub}{\end{subequations}}

\newcommand{\al}{\alpha}
\newcommand{\eps}{\epsilon}

\begin{document}
\title{Phase switching in a voltage-biased Aharonov-Bohm interferometer}
\author{Vadim I. Puller$^1$}
\author{Yigal Meir$^{1,2}$}
\affiliation{$^1$Department of Physics, Ben-Gurion University of
the Negev, Beer Sheva 84105 Israel\\$^{2}$ The Ilse Katz Center
for Meso- and Nano-scale Science and Technology, Ben-Gurion
University, Beer Sheva 84105, Israel}

\begin{abstract}
{Recent  experiment [Sigrist et al., Phys. Rev. Lett. {\bf 98},
036805 (2007)] reported switches between $0$ and $\pi$ in the
phase of Aharonov-Bohm oscillations of the two-terminal
differential conductance through a two-dot ring with increasing
voltage bias. Using a simple model, where one of the dots contains
multiple interacting levels, these findings are explained as a
result of transport through the interferometer being dominated at
different biases by quantum dot levels of different "parity" (i.e.
the sign of the overlap integral between the dot state and the
states in the leads). The redistribution of electron population
between different levels with bias leads to the fact that the
number of switching events is not necessarily equal to the number
of dot levels, in agreement with experiment. For the same reason
switching does not always imply that the parity of levels is
strictly alternating.
Lastly, it is demonstrated that the correlation between the first
switching of the phase and the onset of the inelastic cotunneling,
as well as the sharp (rather than gradual) change of phase when
switching occurs, give reason to think that the present
interpretation of the experiment is preferable to the one based on
electrostatic AB effect.}

\end{abstract}
\date{\today}

 \maketitle

\section{Introduction}
The phase of a transmission coefficient through a quantum dot (QD)
can be experimentally studied using an Aharonov-Bohm (AB)
interferometer with the QD embedded in one of its
arms.\cite{Heiblum,Hackenbroich} Since the early experiments
\cite{Heiblum}, large body of theoretical work has been devoted to
study of the rich phase behavior experimentally discovered. When
studied in two terminal geometry such an interferometer exhibits
"phase rigidity", i.e. the AB oscillations of conductance have
either maximum or minimum at zero magnetic field, which
corresponds to a transmission phase equal to 0 or $\pi$.
Phase rigidity has its origins in time-reversal symmetry (i.e.
unitarity of the transmission matrix). \cite{Buttiker} In "open"
interferometers\cite{Aharony} or two-terminal interferometers in
non equilibrium conditions\cite{Bruder} the phase rigidity may be
broken, i.e. the  phase may change continuously.
Particularly interesting are phenomena connected to "phase
lapses", which are abrupt changes of the transmission phase as a
function of the plunger voltage in the Coulomb blockade valley
(i.e. between two successive Coulomb blockade resonances). While
there has not been yet general agreement about the physics
underlying this phenomenon,\cite{Imry,Gefen,Oreg} it is believed
to be the consequence of the system undergoing transitions between
regimes where transport is dominated by different
levels.\cite{Imry,population_switching}

Recently, an interesting new phase phenomenon has been observed
\cite{Ensslin} in an AB interferometer under non-equilibrium
conditions (i.e. with finite bias applied between the source and
the drain). In this experiment the differential conductance of a
two terminal interferometer was measured as a function of the bias
voltage and magnetic flux through the interferometer. Both arms of
the interferometer contained quantum dots, which were tuned to the
Coulomb blockade regime, thus allowing only for cotunneling
transport. As a function of bias  several switches of the phase of
the AB oscillations  between the values 0 and $\pi$ were observed
with the first switch coinciding with the  onset of inelastic
cotunneling in one of the dots.

A possible explanation for this effect may be  as the result of
electrostatic AB oscillations, as was studied
in Refs.\onlinecite{Nazarov,Wiel}. Some features of the experiment, however,
are inconsistent with such an explanation.
In particular, as mentioned above, there is apparent correlation between the
onset of the phase switching and the onset of
inelastic cotunneling. In addition, the phase of AB oscillation,
which at finite bias is not limited by symmetry
to 0 or $\pi$, is in fact changing abruptly between these values as
a function of bias, whereas in Ref.\onlinecite{Wiel} it changed smoothly.

An alternative explanation (motivated by "population switching"
studies \cite{Imry,population_switching}) is that,
depending on the bias voltage, the elastic contribution to the transport
through the quantum dot may be dominated by different orbital states within
the dot. Depending on the sign of the matrix elements between
the state in the dot and the states in the leads ("parity" of the state),
an electron can acquire phase $0$ or $\pi$, additional to the other phases
acquired when traversing the interferometer. This results in
AB oscillations having maximum or minimum depending on the parity of the
current-transmitting state. Thus, if the two adjacent QD states have
different parity, the phase switching may occur in the regimes when the
state dominating the coherent transport changes from one to the other,
which may happen with increasing bias.

From the simple theory described above one would naively expect that,
if more than two levels in the dot are energetically accessible, multiple
switching events  would occur only when the levels have alternating
parity, and the number of switching events would then be equal to the number
of parity changes. This presumption contradicts the experiment where the
number of switchings can, in fact, be greater than the number of inelastic
onsets,\cite{Ensslin} i.e. greater than the number of the QD levels involved.
In what follows we will show that neither of these expectations is quite true,
and a simple model can reproduce all the features of the
experiment\cite{Ensslin} as result of redistribution of electron population
between QD states. We also  point out that  at finite bias
"population inversion" is not a necessary condition for the phase switching
to be observed.

The plan of the paper is as follows: in section \ref{sec:model} we
first introduce our model of the interferometer, which, after
a Schrieffer-Wolff transformation,\cite{SW} allows to study cotunneling
processes in terms of a single particle
scattered from a local pseudo-spin degree of freedom. In section
\ref{sec:results} this model is solved in terms of rate equations
and the differential conductance is obtained as a function of the
Aharonov-Bohm flux threading the interferometer
and source-drain bias. We compare results with the experiment, showing,
in particular, that the number of switching events can be greater or less
then the number of levels, whose energy spacing is within the bias window.
We discuss how the switching behavior may occur even when the parity of
levels is not strictly alternating, and compare our results with the
electrostatic AB scenario.

\section{Model}
\label{sec:model}
\subsection{Hamiltonian}
\begin{figure}[tbp]
  \includegraphics[width=3.5in]{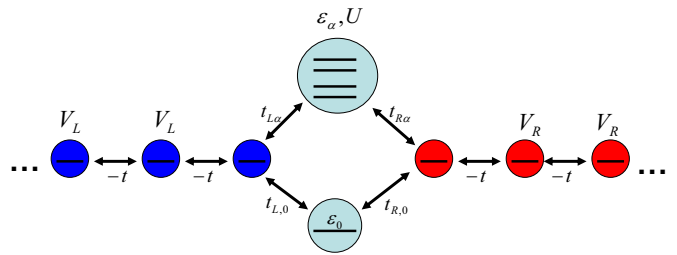}
  \vskip -.6 truecm
  \caption{(color online) Schematic diagram of the modeled device.
  Two infinite tight-binding chains, possibly with different
  chemical potential are coupled to an Aharonov-Bohm interferometer,
  one of the arms of which contains a multi-level quantum dot.
  }\label{fig:loc_interf}
\end{figure}

 We describe the interferometer by a tight-binding
model, schematically shown in Fig. \ref{fig:loc_interf}. The
Coulomb interaction is present in one of the dots, which contains
several electron orbitals (interacting arm), whereas the dot in
other arm of the interferometer (reference arm) has only one
level. The chemical potential is set out of resonance, so that the
transport via the interacting arm occur only by means of
cotunneling processes. The reference arm can be characterized by
its one-particle transmission coefficient, and placing a level in
it is a matter of computational convenience.

In the present experimental context, the Kondo effect is
irrelevant and its effects will be neglected in the present
approach. In addition, for computational simplicity, we assume
that each level carries a different quantum number, which are not
mixed in the leads  (i.e. we disregard difference
between spin and orbital channels). The results obtained below can be shown
to be correct also in the case when the number of channels in the leads is
not equal to the  number of states in the quantum dot (e.g. when all the
levels in the dot are coupled to the same lead channel).
However, since the dot levels in such a system are connected via tunneling
into the leads, the correct calculation is more complicated and requires a
decomposition procedure similar to that developed in
Ref.\onlinecite{Kashcheyevs}.

Cotunneling processes can be described as second order transitions
between different states which contain exactly one electron one in
the dot.\cite{RateEq} By performing a Schrieffer-Wolff transformation\cite{SW}
(Appendix \ref{appx:SW}) we can reduce the problem to that of one particle
scattered by local pseudo-spin degree of freedom. As a result, we can
describe the system by the following (one-particle) Hamiltonian:
\begin{equation} {\cal H} = H_L +H_R + H_D + W_{ref} + W_2 +
W_4 +W_6, \label{eq:H}
\end{equation}
where
\begin{widetext}
\bsub
\bn H_{L}&=&\sum_{\al}\sum_{n=-\infty}^{-1}\left[V_L c_{\al,n}^+ c_{\al,n}-
t\left( c_{\al,n-1}^+ c_{\al,n}+c_{\al,n}^+ c_{\al,n-1}\right)\right], \\
H_{R}&=&\sum_{\al}\sum_{n=1}^{+\infty}\left[V_R c_{\al,n}^+ c_{\al,n}-
t\left(c_{\al,n+1}^+ c_{\al,n}+c_{\al,n}^+ c_{\al,n+1}\right)\right], \\
H_D&=&\sum_{\al}\eps_\al n_\al, \\
W_{ref}&=&-\sum_\al \left( \omega_{rr} c_{\al,1}^+ c_{\al,1}+
\omega_{lr} c_{\al,-1}^+ c_{\al,1}+\omega_{rl} c_{\al,1}^+ c_{\al,-1}+
\omega_{ll} c_{\al,-1}^+ c_{\al,-1}\right),\en
\bn
W_2&=&-\sum_\al \left( v_{rr}^\al c_{\al,1}^+ c_{\al,1}+
v_{lr}^\al c_{\al,-1}^+ c_{\al,1}+v_{rl}^\al c_{\al,1}^+ c_{\al,-1}+
v_{ll}^\al c_{\al,-1}^+ c_{\al,-1}\right),\\
W_4&=&\sum_{\al,\beta\neq\al} \left( v_{rr}^{\beta\al}
c_{\beta,1}^+ d_\al^+d_\beta c_{\al,1}+
v_{lr}^{\beta\al} c_{\beta,-1}^+ d_\al^+d_\beta c_{\al,1}+\right.\\
&&\left.v_{rl}^{\beta\al} c_{\beta,1}^+ d_\al^+d_\beta c_{\al,-1}+
v_{ll}^{\beta\al} c_{\beta,-1}^+ d_\al^+d_\beta c_{\al,-1}\right),\\
W_6&=&-\sum_{\al,\beta\neq\al}
\left(v_{rr}^{\al\al}n_\beta c_{\al,1}^+c_{\al,1}+
v_{lr}^{\al\al}n_\beta c_{\al,-1}^+c_{\al,1}+
v_{rl}^{\al\al}n_\beta c_{\al,1}^+c_{\al,-1}+
v_{ll}^{\al\al}n_\beta c_{\al,-1}^+ c_{\al,-1}\right). \en
\esub
\end{widetext}
with $n_\al=d_\al^+d_\al)$. Here $\al,\beta$ refer to the
different dot levels/lead channels, whereas $n$ is the
tight-binding site label with $n>0 (n<0)$ in the right (left)
lead. $c_{\al,n}^{+} (c_{\al,n})$ creates (annihilates) an
electron in channel $\al$ on site $n$, whereas $d_{\al}^{+}
(d_{\al})$ creates (annihilates) electron in state $\al$  in the
quantum dot with energy $\eps_{\al}$. $V_L, V_R$ are the
potentials applied to the leads, related to the chemical
potentials as $\mu_{L,R}=\eps_F+V_{L,R}$ ($\eps_F$ is the Fermi
energy), $t$ is the hopping integral.

The terms $H_L,H_R$ describe non-interacting leads, $H_D$ is the
Hamiltonian of the dot, $W_{2}$ and $W_{6}$ describe elastic
cotunneling through the quantum dot when it is empty or occupied
by one electron respectively. $W_4$ describes inelastic
cotunneling, i.e. the processes when an electron incident in
channel $\al$ is scattered into channel $\beta$, while the dot
changes its state from $\beta$ to $\al$. $W_{ref}$ describes tunneling
through the reference arm, which we have taken as channel-independent.
The AB phase, $\phi_{AB}$, can be attached to the matrix elements for
the reference arm:
$\omega_{rl}=\omega_{lr}^*=|\omega_{rl}|\exp{(i\phi_{AB})}$.
The explicit dependence of the tunneling parameters $v$ and $\omega$
on the original parameters appear in Appendix \ref{appx:SW}.

\subsection{Scattering matrix}
The energy of the system "QD+incident electron" is conserved, so
 we can start with calculating the scattering matrix, which we
do by solving the Schroedinger equation, \bq
H|\Psi\rangle=E|\Psi\rangle,\eq where $E$ is the energy of the
system "dot+incident electron" for the wave function written as
\bn |\Psi\rangle=\sum_{\al,\beta}\left[\sum_{n=-\infty}^{-1}
\left(A_L^{\al\beta}e^{iq_{\beta} n}+
B_L^{\al\beta}e^{-iq_{\beta} n}\right)d_{\beta}^{+}c_{\al,n}^{+}+
\right.\nonumber \\
\left. \sum_{n=1}^{+\infty}\left(A_R^{\al\beta}e^{ik_\beta n}+
B_R^{\al\beta}e^{-ik_\beta n}\right)d_{\beta}^{+}c_{\al,n}^{+} \right]
\left| 0\right\rangle, \label{eq:wf}\en
where $|0\rangle$ is the state with no particles. The wave vectors
$q_\beta,k_\beta$ are defined as solutions of the equations
\bn E=\eps_\beta+V_L-2t\cos{q_\beta},\nonumber \\
E=\eps_\beta+V_R-2t\cos{k_\beta}\en
($\eps_L(q)=V_L+\eps_q$ and $\eps_R(k)=V_R+\eps_k$,
where $\eps_k=-2t\cos{k}$ are electron tight-binding eigenenergies in the two leads).
Let $M$ be the number of the channels, then
$\mathbf{A}_L,\mathbf{A}_R,\mathbf{B}_L,\mathbf{B}_R$ are vectors with $M^2$
components and the amplitude scattering matrix can be defined as
(more details are given in Appendix \ref{appx:ScattMat})
\bq \left(
      \begin{array}{c}
        \mathbf{B}_L \\
        \mathbf{A}_R \\
      \end{array}
    \right)=\hat{S}
    \left(
      \begin{array}{c}
        \mathbf{A}_L \\
        \mathbf{B}_R \\
      \end{array}
    \right),
    \hat{S}=
    \left(
      \begin{array}{cc}
        \hat{S}_{LL} & \hat{S}_{LR} \\
        \hat{S}_{RL} & \hat{S}_{RR} \\
      \end{array}
    \right).\eq
In order to normalize the wave function and the scattering matrix
to unit incident flux, we multiply $\hat{S}$
 by velocity matrices:
\bn \hat{\mathcal{S}}=\hat{u}^{\frac{1}{2}}\hat{S}\hat{u}^{-\frac{1}{2}},
 \hat{u}=\left(
              \begin{array}{cc}
                \hat{u}_L & 0 \\
                0 & \hat{u}_R \\
              \end{array}
            \right),\en
            where
\bn u_L^{\al,\beta;\al'\beta'}=\delta_{\al,\al'}\delta_{\beta,\beta'}
2t\sin{q_{\beta}},\nonumber \\
u_R^{\al,\beta;\al'\beta'}=\delta_{\al,\al'}\delta_{\beta,\beta'}
2t\sin{k_{\beta}}.\en
\subsection{Transition rates}
In order to get the rate equations we write down the rate at which
electrons incident from channel $\al$ of lead $\mu(=L,R)$ are being
scattered into channel $\al'$ of lead $\mu'$, whereas the impurity
changes its state from $\beta$ to $\beta'$. (In fact, only two types
of processes are possible: a) elastic scattering, when
$\al'=\al,\beta'=\beta$, and b) inelastic scattering, when
$\al'=\beta,\beta'=\al$ ($\al\neq\beta$).)
\begin{widetext}
\bq W_{\mu'\leftarrow\mu}^{\al'\beta'\leftarrow\al\beta}=
\frac{1}{2\pi}\int_{-2t}^{2t}d\eps
\int_{-2t}^{2t}d\eps' f_0(\eps)[1-f_0(\eps')]
\left|\mathcal{S}^{\al'\beta';\al\beta}_{\mu';\mu}
(\eps+V_\mu+\eps_\beta)\right|^2
\delta (\eps+V_\mu+\eps_\beta-\eps'-V_{\mu'}-\eps_{\beta'}).
\label{eq:TransitionRate}\eq
\end{widetext}
Here we wrote explicitly the dependence of the scattering matrix on the
energy of the system "QD+incident electron", whereas
$f_0(\eps)=1/[1+\exp{((\eps-\eps_F)/(k_B T))}]$ is the Fermi distribution
 function; $V_L-V_R=V$ is the bias voltage.

{\em Level occupation numbers.} The resulting rate equations,
describing transitions between different states of the quantum dot
are: \bq
\frac{d}{dt}P_{\beta}(t)=\sum_{\beta'\neq\beta}\left(w^{\beta;\beta'}
P_{\beta'}(t)-w^{\beta';\beta}P_{\beta}(t)\right)\eq
\bq w^{\beta;\beta'}=
\sum_{\mu,\mu'}\sum_{\al,\al'}W_{\mu;\mu'}^{\al\beta;\al'\beta'}=
\sum_{\mu,\mu'}W_{\mu;\mu'}^{\beta\beta';\beta'\beta},\beta\neq\beta',\eq
where $P_{\beta}(t)$ is the probability that the electron
occupying the quantum dot is in the state with energy
$\eps_\beta$. In a stationary state the occupation numbers,
$P_\beta$ can be calculated from equations \bq
\frac{d}{dt}P_{\beta}(t)=0\eq supplemented by the normalization
condition for the total number of electrons occupying the dot. \bq
\sum_{\beta}P_\beta=1.\eq

{\em Current.} The lead-to-lead current is given by
$I=I_{coh}+I_{incoh}$, \bsub \bn
I_{coh}=\sum_{\al,\beta}P_\beta \left(W_{RL}^{\al\beta;\al\beta}-
W_{LR}^{\al\beta;\al\beta}\right),\label{eq:Icoh}\\
I_{incoh}=\sum_{\al,\beta}P_\beta \left(W_{RL}^{\al\beta;\beta\al}-
W_{LR}^{\al\beta;\beta\al}\right).\en\esub
where we have explicitly separated the elastic and inelastic contributions.

The differential conductance is defined as
\bq G(V,\phi_{AB})=\frac{\partial I(V,\phi_{AB})}{\partial V},\eq
where $\phi_{AB}=2\pi\Phi/\Phi_0$ is the Aharonov-Bohm phase
(proportional to the magnetic flux, $\Phi$, divided by the
magnetic flux quantum, $\Phi_0\equiv hc/e$).
\section{Results}
\label{sec:results}
 We provide results obtained from the above
equations for an interferometer that has three levels in the
interacting quantum dot. For the sake of clarity we chose
$\eps_3-\eps_2=2(\eps_2-\eps_1)$, so that inelastic cotunneling
processes involving levels 1 and 2 begin at lower bias than the
cotunneling processes that involve levels 2 and 3. The couplings
of the levels to the leads increase with their energies,
$\Gamma_2/\Gamma_1\simeq 3.2$, $\Gamma_3/\Gamma_1 \simeq 9$. The
complete set of parameters appear in Appendix \ref{appx:par}

\begin{figure}[tbp]
  \includegraphics[width=3.5in]{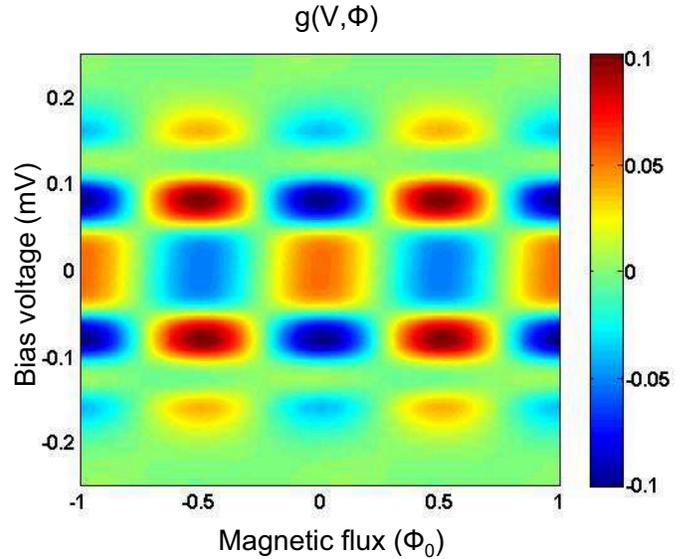}\\
  \caption{(color online) Normalized oscillating part of the differential
  conductance, Eq. (\ref{eq:g}),
   as a function of magnetic flux and bias voltage for the
  set of parameters appearing in Appendix \ref{appx:par}.
  The conductance maximum as a function of flux,
   as zero bias, becomes a minimum and then a maximum again,
   as bias is increased.}\label{fig:G_VF}
\end{figure}

Our main result is Fig. \ref{fig:G_VF}, which shows a plot of the
oscillating component of the differential conductance as a
function of the magnetic flux through the AB ring (horizontal
axis) and the bias voltage applied to the interferometer (vertical
axis). For comparison with the experimental results we normalized
the conductance in the same way as in Ref.\onlinecite{Ensslin} ,
i. e. the quantity shown is \bq
g(V,\phi_{AB})=\frac{G(V,\phi_{AB})-G_{mean}(V)}{G_{mean}(V)}\label{eq:g}\eq
where $G_{mean}(V)$ is the differential conductance averaged over
the period of the magnetic flux, $\phi_{AB}$.
Comparing this figure to Fig. 3 of the experiment by
Sigrist et al.,\cite{Ensslin} we observe the same characteristic features:
(i) depending on the bias voltage, AB oscillations of the differential
conductance may have maximum or minimum at zero magnetic flux;
(ii) the amplitude of the conductance oscillations relative to the
non-oscillating background decreases with increasing bias; and
(iii) the number of switching events (i.e. instances when conductance
AB oscillations change from maximum at $\phi_{AB}=0$ to minimum or vice versa)
is not equal to the number of inelastic onsets (in the case at hand we
have four switchings, but only three inelastic onsets, corresponding
to $\eps_2-\eps_1, \eps_3-\eps_2$ and $\eps_3-\eps_1$, respectively).
Let us discuss these features in greater detail.

\begin{figure}[tbp]
  \includegraphics[width=3.5in]{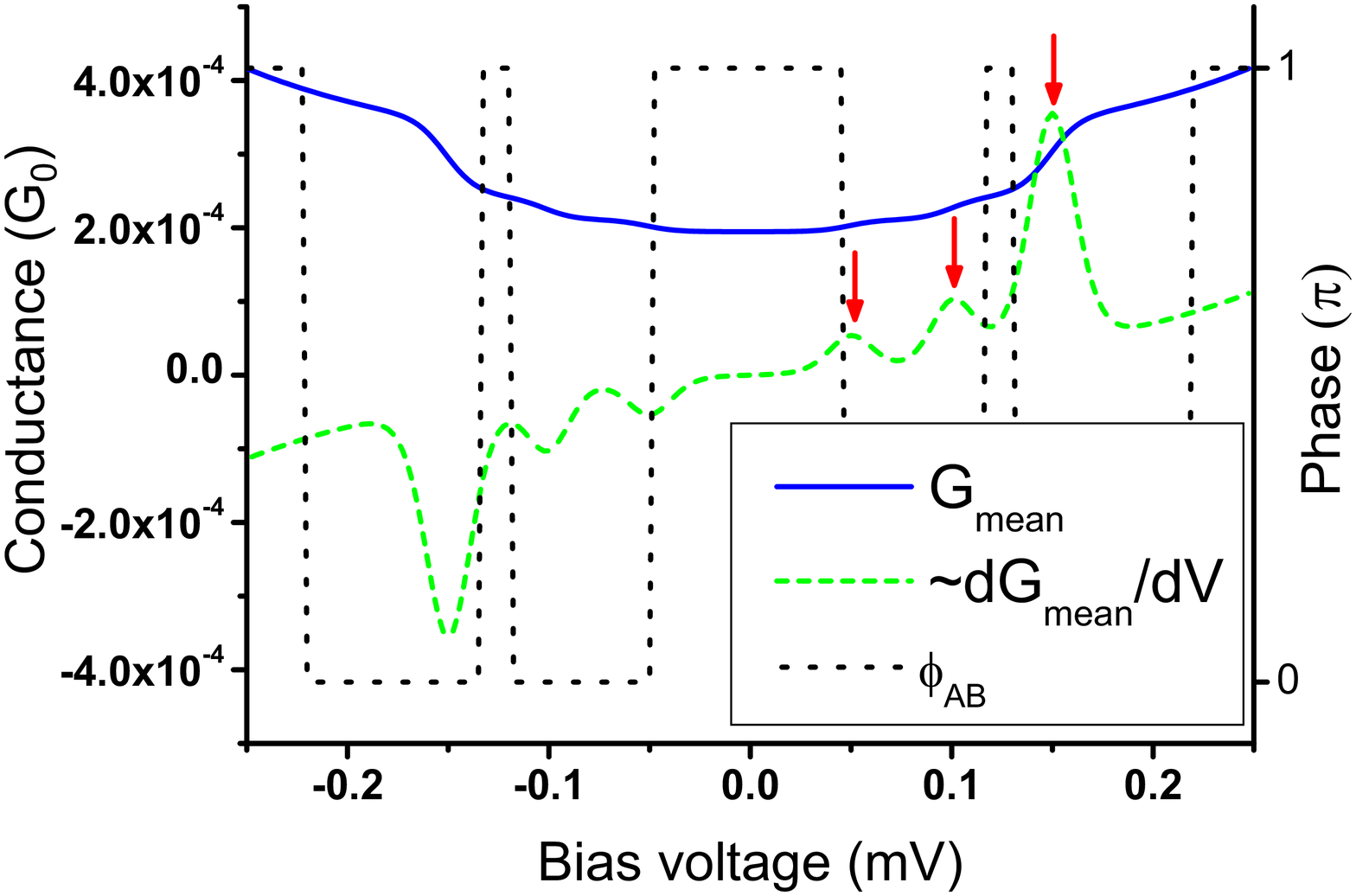}\\
  \caption{(color online) Differential conductance (solid blue),
  its derivative (dash green)
  and AB phase of conductance oscillations as functions of bias
  voltage (dotted black). The onset of the inelastic processes,
   corresponding to $V=\eps_j-\eps_i$ (arrows) are evident.}\label{fig:Gmean}
\end{figure}

\begin{figure}[tbp]
  \includegraphics[width=3.5in]{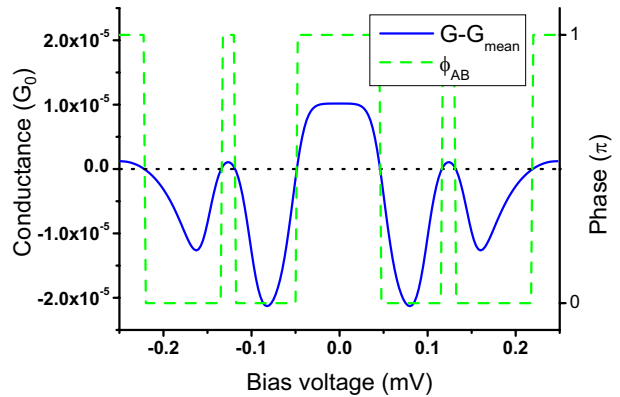}\\
  \caption{(color online) The oscillating contribution to the differential
  conductance at zero field
  (solid blue) and AB phase of conductance
  oscillations (dash green) as a function of bias voltage. }\label{fig:Gosc}
\end{figure}

The mean value and the oscillating component (at zero magnetic
flux) of the differential conductance as functions of bias are
shown in Figures \ref{fig:Gmean} and \ref{fig:Gosc} respectively.
In Fig. \ref{fig:Gmean} one can see clearly the three inelastic
onsets, that is three sharp increases of conductance at bias
voltage equal to spacing between any two of the three levels in
the system (Arrows corresponding to $V=\eps_i-\eps_j$). For clarity
we show on the same plot the first derivative of the conductance and
the phase of conductance oscillations (both multiplied by constant factors).
The phase is defined as the sign of the AB oscillations relative to
$G_{mean}$, i.e. it is $\pi$ if the oscillations have maximum around zero
magnetic flux and $0$ otherwise. It is necessary to point out that at finite
bias the Onsager-B\"{u}ttiker symmetries for a two-terminal structure
do not require that the phase should be restricted to $0,\pi$,\cite{Bruder}
therefore our definition of phase is not mathematically strict.
Nevertheless, for the set of parameters used here (see Fig.~\ref{fig:G_VF})
an extremum is found at zero phase even at finite voltage.


\begin{figure}[tbp]
  \includegraphics[width=3.5in]{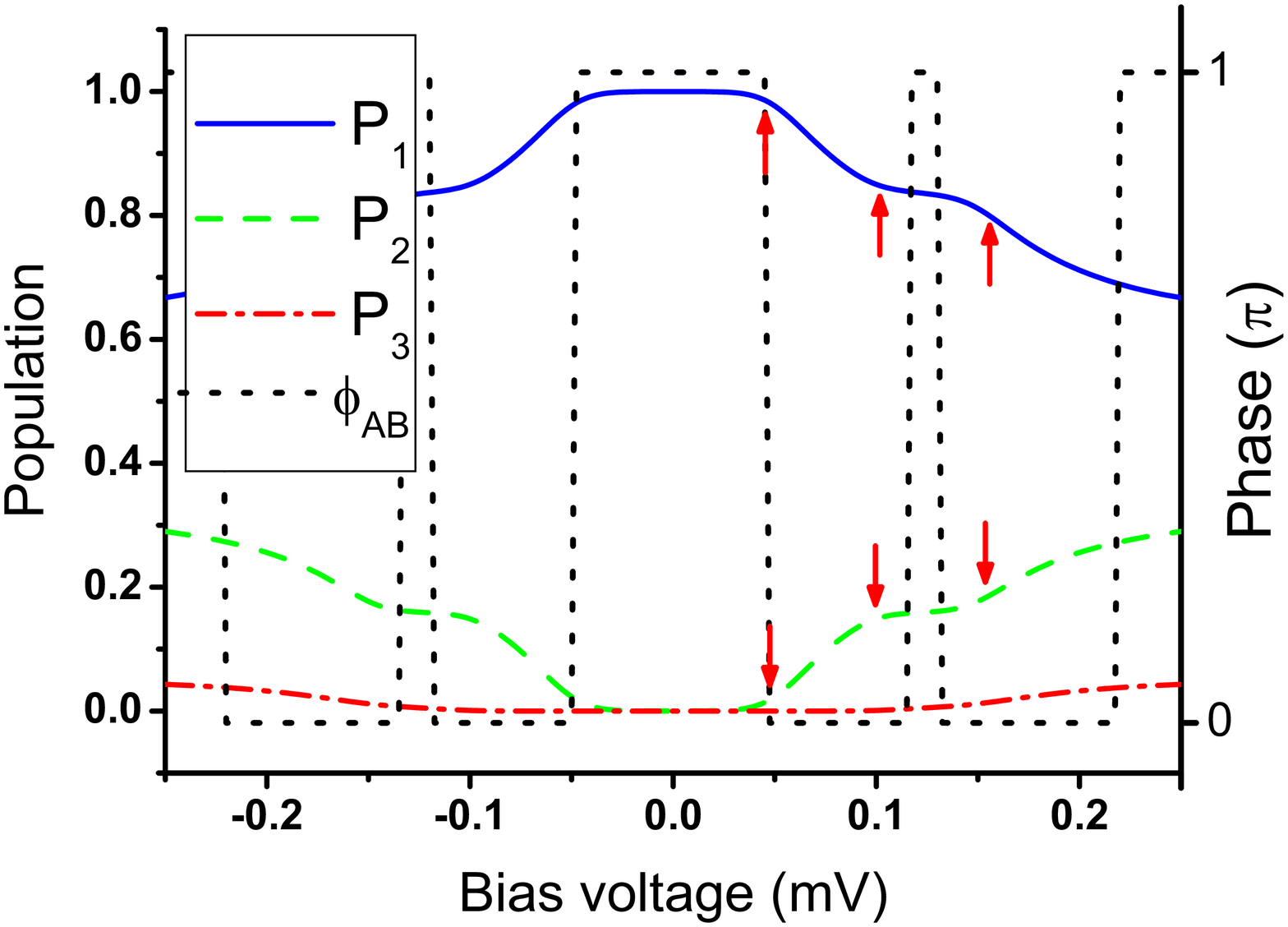}\\
  \caption{(color online) Populations of QD levels as functions of
  bias voltage. With increasing voltage, and with the
  onset of inelastic processes (arrows) there is a significant
  change in the relative occupations of the different
  levels.}\label{fig:Pop}
\end{figure}

The oscillating component of the conductance (Fig. \ref{fig:Gosc})
changes its sign four times. The first change occurs when the bias
reaches value $V=\eps_2-\eps_1$, i.e. when inelastic cotunneling
process becomes possible, in which case an electron occupying
level 1 leaves the dot, whereas another electron can enter the dot
and occupies level 2. While such processes do not contribute to
the coherent AB oscillations (they change the state of the dot,
thus leaving "trace in the environment"\cite{Stern}), they play an
important role in changing the occupation of the different levels
of the dot. In particular, population of level 2, which was for
small bias occupied only due to finite temperature, increases, as
can be seen in Fig. \ref{fig:Pop}, where we show how the
populations of all three levels in the interacting arm of the
interferometer depends on the applied bias (at $\phi_{AB}=0$). AB
oscillations of conductance result from elastic cotunneling
processes, in which electron can tunnel through any of the dot
levels. The relative weights of these processes are  proportional
to the probability for the respective level to be occupied, and
therefore for low bias all tunneling processes are suppressed,
except those that happen via level 1 or via intermediate state in
which the dot is occupied by two electrons (the latter will be
suppressed, if the Coulomb interaction is large). Thus, the
increase of the occupation of level 2 enhances the weight of the
corresponding elastic cotunneling processes, whose contribution to
AB oscillations has phase opposite to that of level 1 (due to the
opposite parity of these two levels). Due to stronger coupling of
level 2 to the leads the oscillations of opposite phase eventually
outweigh those due to level 1 and phase switching occurs.

This interplay between  the level coupling strength and its
population leads to some interesting effects that happen when the
bias is further increased. At the second inelastic onset, i.e.
when $V=\eps_3-\eps_2$, occupation of level 3 starts to grow due
to transitions from level 2. However, the bias is still not high
enough to excite electrons from level 1 to level 3 directly,
therefore level 3 is still being intensively depopulated by
transitions to level 1, which results in simultaneous increasing
of the population of levels 1 and 3, and saturation (or even a
decrease) of the population of level 2. Therefore the second
switching of AB phase occurs due to the change in the relative
populations of levels 1 and 2. It is necessary to stress the role
of the coupling strengths on the population redistribution:  for
low temperatures the population of level 3 may be extremely small,
but the switching still occurs since level 2 is effectively
depopulated by transitions to level 1 via level 3.

Further increase of the bias leads  to increase of the population
of level 2, which again outweighs that of levels 1 and 3 and
causes the third switching. Finally, at the biases greater than
$\eps_3-\eps_1$, direct population of level 3 from the lowest
energy level 1 starts and the contribution of transport with level
3 occupied eventually takes over all other contributions and
results in the fourth switching occurrence.

\begin{figure}[tbp]
  \includegraphics[width=3.5in]{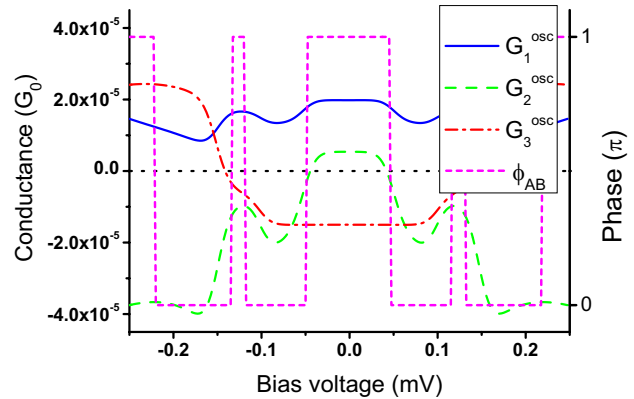}\\
  \caption{(color online) Contributions of different levels in AB
  oscillations as functions of bias at zero magnetic
  flux.}\label{fig:Gosc_par}
\end{figure}

The contributions of  different levels to the AB oscillations are
shown in Fig. \ref{fig:Gosc_par}.
Let us also point out that a contribution of a level to AB
oscillations may not always have the same sign. Thus, there is a
regime in which two levels of different parity (levels 2 and 3)
give the same sign contribution to AB oscillations.  The reason
for that is finite value of Coulomb interaction, which allows for
two cotunneling processes: a) hole tunneling via the occupied
level (with intermediate state being an empty dot), and b)
electron tunneling via an unoccupied level (with intermediate
state being the dot twice occupied). These two processes have
energy denominators of different signs. In other words, the former
process involve exchanging the two electrons and therefore
contributes with the sign different from the latter process. Thus,
for example, the contribution of level 3 to the AB oscillations is
negative as long as this level is unoccupied, but it becomes
positive when the population of this level becomes non-zero. This
change of sign does not occur in the limit of large Coulomb
interaction, when only hole tunneling (or only electron tunneling)
is possible.

\begin{figure}[tbp]
  \includegraphics[width=3.5in]{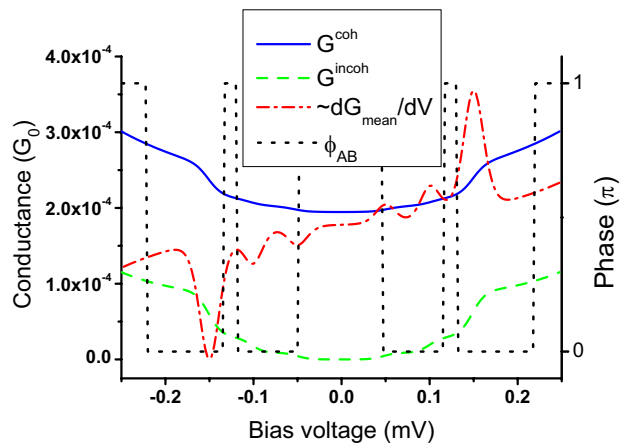}\\
  \caption{(color online) Coherent (elastic cotunneling) and
  incoherent (inelastic cotunneling) contributions to the
  differential conductance (shown respectively in solid blue and
  dashed green).}\label{fig:Gcoh}
\end{figure}

Coherent (elastic cotunneling) and incoherent  (inelastic
cotunneling) contributions to the conductance are depicted in Fig.
\ref{fig:Gcoh}. Inelastic cotunneling never contributes to AB
oscillations and therefore presents a background, reducing
visibility of the oscillations. The contribution due to the
elastic processes contains both oscillating and constant ("elastic
background") components, whose relative size depends on the
relative transmission of the two arms of the interferometer. The
elastic background is minimized when the transmissions through the
two arms of the devices are of the same order.

\begin{figure}[tbp]
  \includegraphics[width=3in]{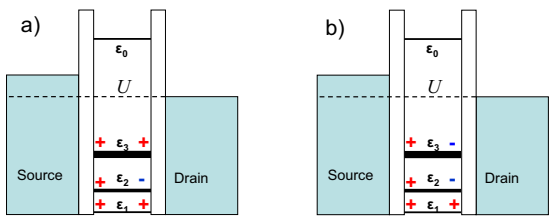}
  \caption{(color online) Energy diagrams for the two cases considered:
  (a) strictly alternating parity of levels, and (b) levels 2 and 3
  having the same parity.}\label{fig:parity}
\end{figure}

\begin{figure}[tbp]
  \includegraphics[width=3.5in]{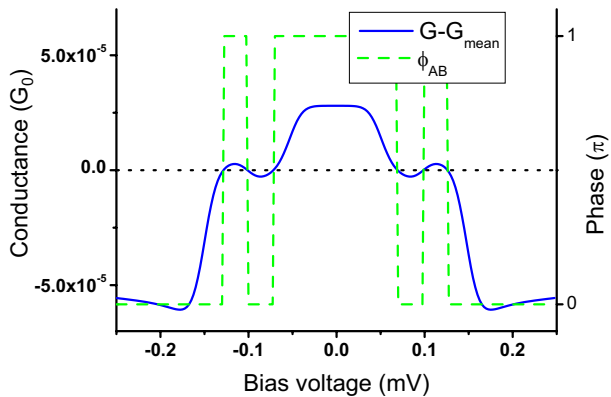}\\
  \caption{(color online) The oscillating contribution to the differential
  conductance (solid blue) and AB phase of conductance
  oscillations (dash green) as functions of bias voltage in the case when
  the parity of levels is not strictly alternating (Fig. \ref{fig:parity}b).
  }\label{fig:Gosc_same_parity}
\end{figure}

So far we have considered the system of levels with strictly
alternating parity, Fig. \ref{fig:parity}a. However, the effect of
redistribution of the dot population between different levels,
considered above, can manifest itself in an interesting way also
in cases when the parity of the levels is not strictly
alternating, such as, e.g., the case shown in Fig.
\ref{fig:parity}b, where level 3 has the same parity as level 2.
The oscillating contribution to the differential conductance of
such a system is shown in Fig. \ref{fig:Gosc_same_parity}, where
the calculations were done with the parameters used above, except
the Coulomb interaction, which was taken to be 2.4 times bigger.
In addition to the first conductance switch (which occur when
inleastic cotunneling with participation of level 2 begins) there
are two more switching evens which appear after the second and
third inelastic onsets (i.e. when level 3 starts being populated
from level 2 and slows populating of the latter, and when level 3
starts to be populated directly from level 1).

Thus, the switching mechanism proposed in this paper has rather moderate
demands in respect to the statistics of level parity in a quantum dot.

One may try to explain the results of Ref. \onlinecite{Ensslin} as
originating from electrostatic AB effect.\cite{Wiel} Indeed, due to
their small transmission, the quantum dots can be thought of as barriers
embedded in the two arms of the interferometer, the whole device thus
being an implementation of the proposal for measuring electrostatic AB
effect made by Nazarov.\cite{Nazarov,Wiel} The main distinction between
the electrostatic AB effect and the scenario described in this paper are
the intervals between switchings. In electrostatic AB effect the switchings
have to occur with a certain bias voltage period, determined by the electron
time-of-flight through the interferometer. This means that switching events
have to be separated by approximately equal bias voltage intervals
(the period varying to the extent that the density-of-states depends on
energy). On the other hand, the mechanism proposed in this paper allows
for the switchings to be separated by unequal bias intervals, as indeed
is the case in the experiment.\cite{Ensslin} Moreover, the first switching
is clearly correlated with the onset of inelastic cotunneling, which cannot
be the case for electrostatic AB effect. Experimentally these two mechanisms
can be compared by studying interferometers with the same size of quantum
dots but different length of the arms, i.e. different period of
electrostatic AB oscillations.

Another important way to distinguish between the two possible explanations
is by observing the characteristic behavior of the phase of AB oscillations
as it changes between $0$ and $\pi$. In electrostatic AB effect this change
is  smooth, almost linear in bias,\cite{Wiel,Ensslin1} whereas in the
experiment by Sigrist et al. the phase is almost constant, except when the
bias is close to its switching value, in which case the phase flows rapidly
between $0$ and $\pi$. The latter phase behavior can be reproduced within the
switching mechanism proposed in this paper. However, it requires application
of many-body scattering theory and cannot be done within the one particle
scattering matrix approach employed above, as it incorrectly describes phase
behavior at small bias. Therefore we leave the discussion of these interesting
phase phenomena to another publication.\cite{PullerMeir}

\section{Conclusions}

We  have addressed the findings of the experiment by Sigrist et
al.\cite{Ensslin} We have shown that switching of the phase of
the AB oscillations of the conductance between $0$ and $\pi$ as a
function of bias can be explained as the result of transport
through the interacting arm of the interferometer being dominated
at different bias by quantum dot states of different "parity".

The  number of switching events is not necessarily equal to the
number of levels in the interacting arm, which agrees with
experiment and finds its explanation in complex redistribution of
electron population between different levels. The same
redistribution implies that conductance switching is not
necessarily the sign of alternating parity of levels within the
quantum dot.

Finally,  the correlation between the first switching of the phase
and the onset of the inelastic cotunneling, as well as unequal separation
of the biases at which the switching occurs, give
reason to think that our explanation of the experiment is
preferable to the one based on electrostatic AB
effect.\cite{Nazarov,Wiel}

\section{Acknowledgement}
The authors would like to thank T. Ihn for sharing with us details
of the experimental work.  We also thank V. Kashcheyevs and T.
Aono for many useful discussions. This work was supported in part
by the ISF and BSF. V.P. is partially supported by Pratt
Fellowship.

\appendix
\begin{widetext}
\section{Schrieffer-Wolff transformation for a quantum dot inserted in
a tight-binding chain}\label{appx:SW}
Let us consider the following Hamiltonian, describing a QD sandwiched
between two leads, where each level of the dot is coupled to a separate
channel in the leads. The channels in the leads are assumed to be
identical, whereas the states in the dot have different energies:
\bq H=H_L+H_R+H_D+H_T \eq
\begin{subequations}
\bn H_D=\sum_{\al}\eps_\al n_\al +\frac{U}{2}
\sum_{\al,\beta\neq\al}n_\al n_\beta, (n_\al=d_\al^+d_\al) \\
H_{L}=-t\sum_{\al,j=-\infty}^{-1}\left(c_{\al,j-1}^+ c_{\al,j}+
c_{\al,j}^+ c_{\al,j-1}\right)+V_L \sum_{\al,j=
-\infty}^{-1} c_{\al,j}^+ c_{\al,j} \\
H_{R}=-t\sum_{\al,j=1}^{+\infty}\left(c_{\al,j+1}^+ c_{\al,j}+
c_{\al,j}^+ c_{\al,j+1}\right)+
V_R \sum_{\al,j=1}^{+\infty} c_{\al,j}^+ c_{\al,j} \\
H_T=\sum_{\al}\left(t_{L\al}d_\al^+c_{\al,-1}+
t_{L\al}^* c_{\al,-1}^+ d_\al +t_{R\al}d_\al^+c_{\al,1}+
t_{R\al}^* c_{\al,1}^+ d_\al\right) \en
\end{subequations}
In order to perform Schrieffer-Wolff transformation \cite{SW}
we first transform the lead Hamiltonians to the representation,
in which they are diagonal:
\bq c_{R\al k}=\sum_{n=1}^{+\infty}\phi_k(n)c_{\al,n}, c_{L\al k}=
\sum_{n=-\infty}^{-1}\phi_k(n)c_{\al,n}, \phi_k(n)=
\sqrt{\frac{2}{\pi}}\sin{(kn)}\eq
The reciprocal relations are
\bq c_{\al,n}=\int_0^\pi dk \phi_k(n)c_{R\al k}, \textrm{ if } n\geq 1;
c_{\al,n}=\int_0^\pi dk \phi_k(n)c_{L\al k}, \textrm{ if } n\leq -1.\eq
The lead Hamiltonians take form:
\bq H_L=\sum_{\al}\int_0^\pi dk (\eps_k+V_L)c_{L\al k}^+ c_{L\al k},
H_R=\sum_{\al}\int_0^\pi dk (\eps_k+V_R)c_{R\al k}^+ c_{R\al k},\eq
where
\bq \eps_k=-2t\cos{(k)}\eq
is the kinetic energy of electron with quantum number $k$.

Now the tunneling Hamiltonian takes form:
\bn H_T=\sum_\al\int_0^\pi dk \left[ t_{L\al}\phi_k(-1)d_\al^+ c_{L\al k}+
t_{L\al}^*\phi_k(-1)c_{L\al k}^+d_\al+ \right.\nonumber \\ \left. t_{R\al}\phi_k(1)d_\al^+ c_{R\al k}+
t_{R\al}^*\phi_k(1)c_{R\al k}^+d_\al\right].\en

In order to perform the SW transformation we need to find an operator $S$
that satisfies the equation
\bq H_T+\left[S,H_L+H_R+H_D\right]=0, \eq
then the transformed Hamiltonian will take form:
\bq \bar{H}=H_L+H_R+H_D+\frac{1}{2}\left[S,H_T\right].\eq
We choose operator $S$ as
\bn S=F-F^+,\nonumber \\
F=\sum_\al \int_0^\pi dk \left[ A_{R\al k}d_\al^+c_{R\al k} +
\sum_{\beta\neq \al}B_{R\al k,\beta}n_\beta d_\al^+ c_{R\al k}+\right.
\nonumber \\ \left.
A_{L\al k}d_\al^+c_{L\al k} +
\sum_{\beta\neq \al}B_{L\al k,\beta}n_\beta d_\al^+ c_{L\al k} \right]\en
This operator is sufficient in the case when the number of channels is two.
For a greater number of channels it is necessary to add terms that account
for the QD being occupied by two, three and more electrons. However, we
neglect such a possibility and consider only single electron entering or
leaving dot. In mathematical terms this is equivalent to performing SW
transformation an a restricted basis, which includes only the states with
zero, one and two electrons in the dot.

Now it is straightforward to show that the coefficients $A,B$ should have forms
\bn A_{R\al k}=\frac{t_{R\al}\phi_k(1)}{\eps_\al-\eps_k-V_R}, A_{L\al k}=\frac{t_{L\al}\phi_k(-1)}{\eps_\al-\eps_k-V_L}, \\
B_{R\al k,\beta}=t_{R\al}\phi_k(1)\left(\frac{1}{\eps_\al+U-\eps_k-V_R}-
\frac{1}{\eps_\al-\eps_k-V_R}\right),\nonumber \\ B_{L\al k,\beta}=t_{L\al}\phi_k(-1)\left(\frac{1}{\eps_\al+U-\eps_k-V_L}-
\frac{1}{\eps_\al-\eps_k-V_L}\right). \nonumber\en
The transformed Hamiltonian is
\bq \bar{H}=H_L+H_R+H_D+\sum_{j=1}^{6}W_j, \eq
where
\bn W_1=\frac{1}{2}\sum_{\al}\int_0^\pi dk \left[ A_{R\al k}t_{R\al}^*
\phi_k(1)+ A_{L\al k}t_{L\al}^*\phi_k(-1)+\right. \nonumber \\ \left.
 A_{R\al k}^*t_{R\al}\phi_k(1)+
A_{L\al k}^*t_{L\al}\phi_k(-1)\right)n_\al=\sum_{\al}\Delta\eps_\al n_\al \en
describes the correction to the dot energies due to coupling to the leads.
In the following we will neglect this term - the appropriate corrections
can be included in energies $\eps_\al$ of the Hamiltonian $H_D$.
\bn W_2=-\sum_\al \left( v_{rr}^\al c_{\al,1}^+ c_{\al,1}+
v_{lr}^\al c_{\al,-1}^+ c_{\al,1}+v_{rl}^\al c_{\al,1}^+ c_{\al,-1}+
v_{ll}^\al c_{\al,-1}^+ c_{\al,-1}\right)\\
v_{rr}^\al \approx \frac{|t_{R\al}|^2}{\eps_\al-\eps_F-V_R} ,
v_{ll}^\al \approx \frac{|t_{L\al}|^2}{\eps_\al-\eps_F-V_L} ,\nonumber \\
v_{lr}^\al=\frac{1}{2}t_{R\al}t_{L\al}^* \left(\frac{1}{\eps_\al-\eps_F-V_R}+
\frac{1}{\eps_\al-\eps_F-V_L}\right),
v_{rl}^\al=(v_{lr}^\al)^* .\nonumber\en
$W_2$ describes elastic cotunneling through the QD without account for
interactions. Following SW we replaced in the denominators electron
kinetic energy, $\eps_k$, by the Fermi energy. It is essential to keep
in mind that, since the leads are kept at different bias, their chemical
potentials are different: $\mu_L=\eps_F+V_L,\mu_R=\eps_F+V_R$.
\bn W_3=\frac{1}{2}\sum_{\al,\beta\neq \al}\int_0^\pi dk \int_0^\pi dk'
\left[ B_{R\al k,\beta}t_{R\beta}\phi_{k'}(1)
d_\beta^+d_\al^+c_{R\al k}c_{R\beta k'}+\right. \nonumber \\ \left.
B_{R\al k,\beta}t_{L\beta}\phi_{k'}(-1)
d_\beta^+d_\al^+c_{R\al k}c_{L\beta k'}+
B_{L\al k,\beta}t_{R\beta}\phi_{k'}(1)d_\beta^+d_\al^+c_{L\al k}c_{R\beta k'}+
\right. \nonumber \\ \left.
B_{L\al k,\beta}t_{L\beta}\phi_{k'}(1)d_\beta^+d_\al^+c_{L\al k}c_{L\beta k'}+
H.c. \right]\en
$W_3$ describes simultaneous tunneling of two electrons into or out of the dot.
In the following we neglect this term, since the corresponding processes are
energetically unavailable.
\bn W_4=\sum_{\al,\beta\neq\al}\left[ v_{rr}^{\beta\al}
c_{\beta,1}^+ d_\al^+d_\beta c_{\al,1}+
v_{lr}^{\beta\al} c_{\beta,-1}^+ d_\al^+d_\beta c_{\al,1}+
\right. \nonumber \\ \left.
v_{rl}^{\beta\al} c_{\beta,1}^+ d_\al^+d_\beta c_{\al,-1}+
\right. \nonumber \\ \left.
v_{ll}^{\beta\al} c_{\beta,-1}^+ d_\al^+d_\beta c_{\al,-1}\right],\\
v_{rr}^{\beta\al}=\frac{1}{2}t_{R\al}t_{R\beta}^*
\left(\frac{1}{\eps_\al+U-\eps_F-V_R}-\frac{1}{\eps_\al-\eps_F-V_R}+
\right. \nonumber \\ \left.
\frac{1}{\eps_\beta+U-\eps_F-V_R}-\frac{1}{\eps_\beta-\eps_F-V_R}\right)
\nonumber \\
v_{ll}^{\beta\al}=\frac{1}{2}t_{L\al}t_{L\beta}^*
\left(\frac{1}{\eps_\al+U-\eps_F-V_L}-\frac{1}{\eps_\al-\eps_F-V_L}+
\right. \nonumber \\ \left.
\frac{1}{\eps_\beta+U-\eps_F-V_L}-\frac{1}{\eps_\beta-\eps_F-V_L}\right)
\nonumber \\
v_{rl}^{\beta\al}=\frac{1}{2}t_{L\al}t_{R\beta}^*
\left(\frac{1}{\eps_\al+U-\eps_F-V_L}-\frac{1}{\eps_\al-\eps_F-V_L}+
\right. \nonumber \\ \left.
\frac{1}{\eps_\beta+U-\eps_F-V_R}-\frac{1}{\eps_\beta-\eps_F-V_R}\right)
\nonumber \\
v_{lr}^{\beta\al}=(v_{rl}^{\al\beta})^* \nonumber\en
$W_4$ describes inelastic cotunneling through the quantum dot, i.e.
the tunneling events when the dot changes its state, while the
incident electron is transferred from one lead to the other or reflected back.
\bn W_5=\frac{1}{2}\sum_{\al,\beta\neq\al}\int_0^\pi dk \left(B_{R\al k,\beta}
t_{R\al}^*\phi_k(1)+B_{L\al k,\beta}t_{L\al}^*\phi_k(-1)+C.c.\right)
n_\al n_\beta =\nonumber \\ \frac{1}{2}\sum_{\al,\beta\neq \al}\Delta
U_{\al\beta}n_{\al}n_{\beta}\en
$W_5$ describes correction to the Coulomb energy and is expected to be small;
in the following we neglect this term.
Finally, we also have
\bn W_6=-\sum_{\al,\beta\neq\al}\left(v_{rr}^{\al\al}n_\beta
c_{\al,1}^+c_{\al,1}+v_{lr}^{\al\al}n_\beta c_{\al,-1}^+c_{\al,1}+
v_{rl}^{\al\al}n_\beta c_{\al,1}^+c_{\al,-1}+
v_{ll}^{\al\al}n_\beta c_{\al,-1}^+ c_{\al,-1}\right)\\
v_{rr}^{\al\al}=|t_{R\al}|^2\left(\frac{1}{\eps_\al+U-\eps_F-V_R}-\frac{1}
{\eps_\al-\eps_F-V_R}\right),\nonumber \\
v_{ll}^{\al\al}=|t_{L\al}|^2\left(\frac{1}{\eps_\al+U-\eps_F-V_L}-
\frac{1}{\eps_\al-\eps_F-V_L}\right),\nonumber \\
v_{lr}^{\al\al}=\frac{1}{2}t_{R\al}t_{L\al}^*
\left(\frac{1}{\eps_\al+U-\eps_F-V_R}-\frac{1}{\eps_\al-\eps_F-V_R}+
\right. \nonumber \\
\left. \frac{1}{\eps_\al+U-\eps_F-V_L}-\frac{1}{\eps_\al-\eps_F-V_L}\right)
\nonumber \\
v_{rl}^{\al\al}=(v_{lr}^{\al\al})^*,\nonumber
\en
which describes the correction to elastic cotunneling contribution, $W_2$,
which arises when the dot is occupied by an electron ($W_2$ is elastic
cotunneling through an empty dot: "hole tunneling").

For the reference arm, in which we neglect the Coulomb interaction,
the Schrieffer-Wolff transformation readily produces the only tunneling term:
\bn W_{ref}=-\sum_\al \left( \omega_{rr} c_{\al,1}^+ c_{\al,1}+
\omega_{lr} c_{\al,-1}^+ c_{\al,1}+\omega_{rl} c_{\al,1}^+ c_{\al,-1}+
\omega_{ll} c_{\al,-1}^+ c_{\al,-1}\right)\\
\omega_{rr} \approx \frac{|t_{R,0}|^2}{\eps_0-\eps_F-V_R} ,
\omega_{ll} \approx \frac{|t_{L,0}|^2}{\eps_0-\eps_F-V_L} ,\nonumber \\
\omega_{lr}=\frac{1}{2}t_{R,0}t_{L,0}^* \left(\frac{1}{\eps_0-\eps_F-V_R}+
\frac{1}{\eps_0-\eps_F-V_L}\right),
\omega_{rl}=(\omega_{lr})^* ,\nonumber\en
where $\eps_0$ and $t_{L,0},t_{R,0}$ are the level energy and the tunneling
amplitudes for this arm, which we assume to be independent on the channel
index.

We neglect $W_{1,5}$ describing corrections to the coefficients and
$W_3$ describing two-particle tunneling, but keep the terms $W_{2,4,6}$
which, together with $H_L,H_R,H_D$ and $W_{ref}$ form Hamiltonian
${\cal H}$, Eq. (\ref{eq:H}), used in this paper.

\section{Scattering matrix}\label{appx:ScattMat}
Using the scattering wave function in the form of Eq.
(\ref{eq:wf}), it is easy to show that the amplitudes satisfy
equations \bn \hat{\mathcal{M}} \left(
      \begin{array}{c}
        \mathbf{B}_L \\
        \mathbf{A}_R \\
      \end{array}
    \right)=-
    \hat{\mathcal{N}}\left(
      \begin{array}{c}
        \mathbf{A}_L \\
        \mathbf{B}_R \\
      \end{array}
    \right),\Rightarrow \hat{S}=-\hat{\mathcal{M}}^{-1}\hat{\mathcal{N}};\\
    \hat{\mathcal{M}}=\left(
                        \begin{array}{cc}
                          \hat{\mathcal{M}}_{LL} & \hat{\mathcal{M}}_{LR} \\
                          \hat{\mathcal{M}}_{RL} & \hat{\mathcal{M}}_{RR} \\
                        \end{array}
                      \right),
\hat{\mathcal{N}}=\left(
                        \begin{array}{cc}
                          \hat{\mathcal{N}}_{LL} & \hat{\mathcal{N}}_{LR} \\
                          \hat{\mathcal{N}}_{RL} & \hat{\mathcal{N}}_{RR} \\
                        \end{array}
                      \right).
    \en
The only non-zero elements of matrices $\hat{\mathcal{M}},\hat{\mathcal{N}}$
are
\bn \mathcal{M}_{LL}^{\al\beta;\al\beta}=
t-\left(v_{ll}^\al+(1-\delta_{\al,\beta})v_{ll}^{\al\al}\right)e^{iq_\beta},
\mathcal{M}_{LL}^{\al\beta;\beta\al}=(1-\delta_{\al,\beta})
v_{ll}^{\al\beta}e^{iq_\al},\nonumber \\
\mathcal{M}_{LR}^{\al\beta;\al\beta}=
-\left(v_{lr}^\al+(1-\delta_{\al,\beta})v_{ll}^{\al\al}\right)e^{ik_\beta},
\mathcal{M}_{LR}^{\al\beta;\beta\al}=(1-\delta_{\al,\beta})
v_{lr}^{\al\beta}e^{ik_\al},\nonumber \\
\mathcal{M}_{RL}^{\al\beta;\al\beta}=
-\left(v_{rl}^\al+(1-\delta_{\al,\beta})v_{rl}^{\al\al}\right)e^{iq_{\beta}},
\mathcal{M}_{RL}^{\al\beta;\beta\al}=(1-\delta_{\al,\beta})v_{rl}^{\al\beta}
e^{iq_\al},\nonumber \\
\mathcal{M}_{RR}^{\al\beta;\al\beta}=t-\left(v_{rr}^\al+(1-\delta_{\al,\beta})
v_{rr}^{\al\al}\right)e^{ik_\beta},
\mathcal{M}_{RR}^{\al\beta;\beta\al}=(1-\delta_{\al,\beta})v_{rr}^{\al\beta}
e^{ik_\al},
\en
and
\bn \mathcal{N}_{LL}^{\al\beta;\al\beta}=
t-\left(v_{ll}^\al+(1-\delta_{\al,\beta})v_{ll}^{\al\al}\right)e^{-iq_\beta},
\mathcal{N}_{LL}^{\al\beta;\beta\al}=(1-\delta_{\al,\beta})v_{ll}^{\al\beta}
e^{-iq_\al},\nonumber \\
\mathcal{N}_{LR}^{\al\beta;\al\beta}=-\left(v_{lr}^\al+(1-\delta_{\al,\beta})
v_{ll}^{\al\al}\right)e^{-ik_\beta},
\mathcal{N}_{LR}^{\al\beta;\beta\al}=(1-\delta_{\al,\beta})v_{lr}^{\al\beta}
e^{-ik_\al},\nonumber \\
\mathcal{N}_{RL}^{\al\beta;\al\beta}=-\left(v_{rl}^\al+(1-\delta_{\al,\beta})
v_{rl}^{\al\al}\right)e^{-iq_{\beta}},
\mathcal{N}_{RL}^{\al\beta;\beta\al}=(1-\delta_{\al,\beta})v_{rl}^{\al\beta}
e^{-iq_\al},\nonumber \\
\mathcal{N}_{RR}^{\al\beta;\al\beta}=t-\left(v_{rr}^\al+(1-\delta_{\al,\beta})
v_{rr}^{\al\al}\right)e^{-ik_\beta},
\mathcal{N}_{RR}^{\al\beta;\beta\al}=(1-\delta_{\al,\beta})v_{rr}^{\al\beta}
e^{-ik_\al}.
\en
\end{widetext}

\section{Parameters used in the Results Section}\label{appx:par}
 The level energies were taken to be $\eps_1=-.4
meV,\eps_2=-.35 meV,\eps_3=-.25 meV$; the tunneling amplitudes
are: $t_{L,1}=50\mu eV, t_{L,2}=88.9\mu eV, t_{L,3}=150\mu eV$,
$t_{L,1}=50\mu eV, t_{L,2}=-88.9\mu eV,
 t_{L,3}=150\mu eV$, which correspond to the level widths:
$\Gamma_{L,1}=\Gamma_{R,1}=.25\mu eV, \Gamma_{L,2}=\Gamma_{R,2}=.79\mu eV,
\Gamma_{L,3}=\Gamma_{R,3}=2.25\mu eV$ (in tight-bonding picture level
width is defined as $\Gamma=|t_{\mu,\al}|^2/t$). The Coulomb interaction is
$U=5 meV$. The parameters of the reference arm are: $\eps_0=-1meV$,
$t_{L,0}=t_{R,0}=200\mu eV$, $\Gamma_{L,0}=\Gamma_{L,0}=4\mu eV$.
The temperature is taken to be $T=60 mK$.

\end{document}